\documentclass[aps,prl,twocolumn,groupedaddress,showpacs]{revtex4}
\usepackage[latin1]{inputenc}
\usepackage[dvips,final]{epsfig}
\usepackage{graphicx,amsmath}

\begin{document}

\title{Sound focusing by gradient index sonic lenses}

\author{Alfonso Climente}
\author{Daniel Torrent}
\author{Jos\'e S\'anchez-Dehesa}
\email[]{jsdehesa@upvnet.upv.es}
\affiliation{Wave Phenomena Group, Departamento de Ingenier\'{\i}a Electr\'onica, Universidad Polit\'ecnica de Valencia, Camino de Vera s.n., E-46022 Valencia, Spain}


\date{\today}
\begin{abstract}
Gradient index sonic lenses based on two-dimensional sonic crystals are here designed, fabricated and characterized. 
The index-gradient is achieved in these type of flat lenses by a gradual modification of the sonic crystal filling fraction along the direction perpendicular to the lens axis. 
The focusing performance is well described by an analytical model based on ray theory as well as by numerical simulations based on the multiple-scattering theory. 
\end{abstract}

\maketitle
A two-dimensional (2D) sonic crystal (SC) is just a periodic distribution of solid cylinders in air with their axis parallel aligned. 
The properties of these structures have been already analyzed in a broad range of frequencies. Thus, for frequencies of the order of the lattice separation they present a band of frequencies (bandgap) in which the sound propagation is forbidden because of Bragg reflection\cite{San98-gaps}. 
However, in the range of low frequencies (homogenization limit) they behave like homogeneous media whose effective acoustic parameters, dynamical mass density and bulk modulus, basically depend on the lattice filling fraction\cite{Kro03,Tor06-Hom,TorPRB06}.  

The homogenization properties of SCs have been employed to design refractive devices like, for example, acoustic lenses whose focusing properties are based on their external curved surfaces\cite{MeyerBook,Cer02-lens,Gup03} or Fabry-Perot like acoustic interferometers\cite{SanPRB03}.
More recently, gradient index (GRIN) acoustic lenses based on SCs have been proposed\cite{TorNJP07,LinPRB09}. Like their optical counterparts, the proposed GRIN SC lenses have flat surfaces and their index gradient is obtained by changing the SC filling fraction, which is directly related to the local index of refraction. An advantage of GRIN lenses in comparison with curved lenses is their easy fabrication. The bending of sound waves obtained by GRIN acoustics structures have been also proposed to produce acoustic mirages in a lab\cite{Chin09}.

In this letter 2D GRIN SC lenses have been fabricated and characterized by an specifically designed experimental set up. 
Each lens is made of a rectangular cluster of aluminum cylinders distributed in an hexagonal lattice. 
Their focusing properties are here comprehensively studied either experimentally as well as theoretically. 
We demonstrate that their focusing behavior is a true GRIN effect that is fairly well reproduced by simulations based in the multiple scattering theory and it is also supported by an analytical model based on ray theory. 
To the best of our knowledge, no experimental demonstration of GRIN SC lenses had been presented up to date.

\begin{figure}
\includegraphics[height=50mm,width=\columnwidth]{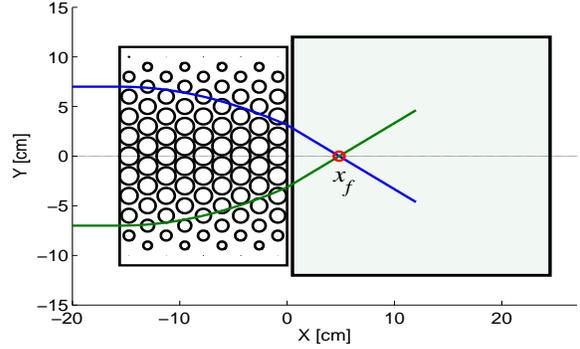}
\caption{\label{fig:FigMEsq}
Schematic view of a GRIN lens made of nine columns of metal rods. Circles define the rods sections, which are position dependent. The sound impinging the lens from the left bends towards the $x$ axis and focuses at position $x_f$ (red dot). The gray rectangle defines the area where the data are acquired.}
\end{figure}

The acoustical axis of GRIN lenses considered here is defined along the $x-$axis and, therefore, the refractive index $n(y)$ changes along the $y-$axis. The local variation of $n$ is obtained by changing the cylinders radii as explained in Ref. \cite{TorNJP07}. A GRIN lens and the region of data acquisition are schematically shown in Fig. \ref{fig:FigMEsq}. 
\par 
The index profile $n(y)$ is designed with the goal of having low aberration of the focal spot. We choose a hyperbolic secant profile because of its demonstrated property of focusing sound inside large enough GRIN structures without aberration \cite{LinPRB09}:
\begin{equation}
\label{eq:RefIndex}
n(y)=n_{0}\ \text{sech}(\alpha y),
\end{equation}	
where $\alpha$ is a constant defined as
\begin{equation}
\alpha=\frac{1}{h}\text{cosh}^{-1}(\frac{n_{0}}{n_{h}}),
\end{equation}
where $h$ is the half-height of the lens, $n_{0}$ is the refractive index on the $x-$axis ($y=$0) and $n_{h}$ that at the lens edges ($y=\pm$ h).
\par 
The differential equation satisfied by the energy rays can be obtained\cite{GomBook02} and by solving it, the position of the focal spot outside the lens, $x_f$, is obtained 
\begin{equation}
\label{eq:FocalDistanceExt}
x_f=d-\frac{y(d)}{y'(d)}\sqrt{\frac{1-(y'(d))^2\left[n^2(y(d))-1\right]}{n^2(y(d))}},
\end{equation}
where $d$ is the lens thickness.
\par
 A 2D GRIN SC can be considered as a homogenized medium made of subwavelength discrete units. The units are metal cylinders that are distributed in rows of equal radii. A given row $\ell$ is parallel to the $x-$axis and it is independently homogeneous and isotropic. The refractive index at row with coordinate $y_\ell$ is adjusted by taking into account that, in the low frequency limit, a cluster of rigid cylinders embedded in a fluid or gas background behaves like a homogeneous medium whose effective refractive index $n_{eff}$ mainly depends on the fraction of volume occupied by the cylinder in the SC or filling fraction\cite{Tor06-Hom,TorPRB06}.

\par 
The 2D GRIN SC lenses studied are designed with a profile $n(y)$ as closer as possible to that in Eq. \eqref{eq:RefIndex} with  $n_{0}=1.337$ and $n_{h}=1$.  
Each lens is made of a rectangular cluster of aluminum cylinders distributed in an hexagonal lattice with parameter $a=2cm$ (see Fig. \ref{fig:FigMEsq}). 
The radiii $R_{\ell}$ of cylinders at row  $y_\ell$ is obtained by imposing the condition $n_{eff}(f)=n(y_\ell)$, where $f=\frac{2\pi}{\sqrt{3}}(R_{\ell}/a)^2$ is the filling fraction of the hexagonal lattice.
Table \ref{tab:TabRad} reports the values of $R_{\ell}$ obtained. 
Note that values corresponding to the lower half of the SC lens are the only ones reported since the lens is symmetric with respect to the $x-$axis. 
Odd layers have 11 cylinders while even layers have 10 cylinders. 
Column $n(y_\ell)$ in Table report the exact hyperbolic secant profile while those under $n_{eff}(y_\ell)$ are those effectively achieved in samples. 

\par 

\begin{table}
\caption{\label{tab:TabRad} Effective refractive index profile $n_{eff}(y)$ of the 2D GRIN SC lenses studied here.}
\begin{ruledtabular}
\begin{tabular}{c c c c c c c}
& & &\multicolumn{2}{c}{Odd layers} & \multicolumn{2}{c}{Even layers}\\
\hline
$y_{\ell}$(mm) &$n(y_\ell)$& $n_{eff}(y_\ell)$ & $\ell$& $R_\ell$(mm) &$\ell$& $R_\ell$(mm)\\
\hline
	-100 & 1     & 1     & 1  & 0     & - & -   \\
  -90  & 1.052 & 1.070 & -  & -     & 1 & 8.0 \\
	-80  & 1.102 & 1.107 & 2  & 10.0  & - & -   \\
	-70  & 1.149 & 1.151 & -  & -     & 2 & 12.0 \\
	-60  & 1.193 & 1.201 & 3  & 14.0  & - & -   \\
	-50  & 1.232 & 1.228 & -  & -     & 3 & 15.0\\
	-40  & 1.266 & 1.258 & 4  & 16.0  & - & -   \\		
	-30  & 1.293 & 1.293 & -  & -     & 4 & 17.0\\	
	-20  & 1.314 & 1.313 & 5  & 17.5  &   & - \\	
	-10  & 1.326 & 1.339 & -  & -     & 5 & 18.0 \\
	  0  & 1.330 & 1.339 & 6  & 18.0  & - & -  \\
 \end{tabular}
\end{ruledtabular}
\end{table}
\par 

Let us point out that the chosen lattice parameter determines an upper frequency limit for the homogenization. Since the homogenization condition is accomplished while $\lambda\geq 4a$\cite{Tor06-Hom}, where $a=$2 cm, for frequencies $\nu\geq 4.2kHz$ the homogenization condition is broken.
\par 
\par 
The lenses have been characterized in a 2D anechoic chamber specifically designed for the parameters of the lenses. 
The chamber consists of two parallel metal sheets separated a distance $H=5cm$. 
For the frequencies analyzed here, from 3.5 kHz up to 4.5 kHz, there are two modes propagating trough the chamber, but the second one has a much larger wavelength and is considered as an offset. 
The sound waves are generated with a UDE AC-150 column loudspeaker separated 2 meters from the chamber in order to get approximately a plane wavefront at the entrance. 
The input signal is a Gaussian pulse centered at the frequencies in between the range under study. A total of eleven frequencies are comprehensively analyzed; i.e., from 3.5 kHz till 4.5 kHz in steps of 0.1 kHz.
Measurements made without lens confirm that the propagating field inside the chamber has practically a plane wavefront. 
We use a B\&K 4958 microphone to record the transmitted sound behind the lens and use the PCI Acquisition Card NI-5105 for processing the data in a computer. 
Data are acquired in a square area of 24$\times$24 cm$^2$, which is represented in Fig. \ref{fig:FigMEsq}. The acquisition points are separated by equal distance along the $x-$ and $y-$axis; i.e., $\Delta x =\Delta y=$ 2cm.

Seven samples have been experimentally characterized. The lenses differ in the number $N$ of layers defining their thickness along the $x-$axis; from 4 to 10. 
The corresponding pressure maps have been measured and compared with those obtained by numerical simulations based on the multiple scattering theory. Besides, the focal spots $x_f$ have been also located and their values are compared with those predicted by Eq. \eqref{eq:FocalDistanceExt}.  Measurements and simulations are performed in the range of frequencies explored every 0.1 kHz. 
\par 
Numerical simulations are performed by using a 2D multiple scattering algorithm as described in Refs. \onlinecite{Gup03,SanPRB03}, where the reader is addressed for details.
 In brief, a 2D sound wave with a plane wavefront impinges the lens from the left and the pressure map is calculated in the opposite side. 
\begin{figure}
\includegraphics[width=75mm]{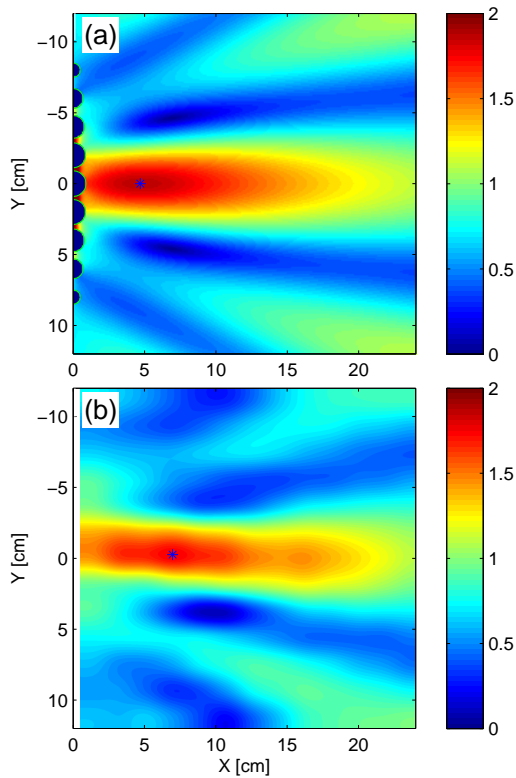}[t]
\caption{\label{fig:FigPF45-9}(a) Sound amplification map generated by 4.5 kHz sound waves impinging a 9 layers thick 2D GRIN SC lens. 
(b) The corresponding map obtained by using a multiple scattering algorithm. Maps are obtained in an area 24$\times$24 $cm^2$ near the lens (see Fig. \ref{fig:FigMEsq}). 
The asterisks mark the focal spot $x_f$.}
\end{figure}
\begin{figure}
\includegraphics[width=75mm]{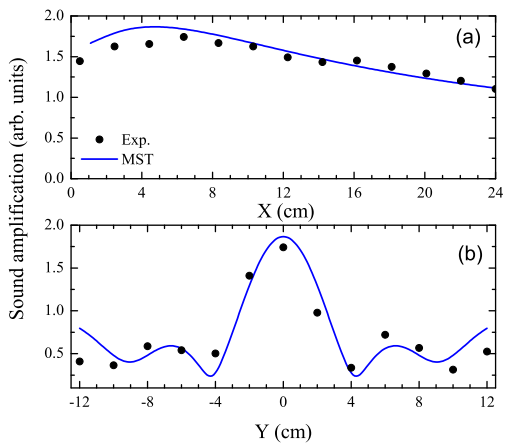}
\caption{\label{fig:Cut45-9} (a) Longitudinal and (b) transversal cuts extracted from the maps shown in Fig. \ref{fig:FigPF45-9}. Simulations based on multiple scattering theory (blue curves) and experimental data (black dots) are depicted.}
\end{figure}
As a typical example of results obtained, Figs. \ref{fig:FigPF45-9}(a) and \ref{fig:FigPF45-9}(b) show the sound amplification maps for a 9 layer thick GRIN lens for the frequency of 4.5 kHz; their corresponding wavelength, $\lambda\approx$ 3.8 $a$, roughly corresponds to the cutoff defining the homogenization limit,
The focusing effects, focal spot and diffraction lobes, are clearly seen in both maps, measured [see Fig. \ref{fig:FigPF45-9}(a)] and calculated [see \ref{fig:FigPF45-9}(b)].
 A better comparison between data and simulations is given in Figs. \ref{fig:Cut45-9}(a) and  \ref{fig:Cut45-9}(b) that show the longitudinal and transversal profiles passing through the focal spot obtained from the previous pressure maps. Note the good agreement exists between data and simulations. A similar agreement is also obtained for the other frequencies analyzed.
\par
In spite of the agreement reported above between theory and experiment a comparison with the position of the focal spot $x_f$ predicted by the ray theory of Eqs. (1) to (3) has been also performed in order to support that sound focusing is a truly refractive effect produced by the bending of waves inside the GRIN lens. The parameters used in Eq. \eqref{eq:FocalDistanceExt} are determined as follows. 
The lens half-height $h$ is obtained from $2h=N_ya$ where $N_y$ is the maximum number of cylinders along the $y-$axis, $N_y=$11.
The lens thickness $d$ is obtained from $d=N\frac{\sqrt{3}}{2}a+k$, where $N$ is the number of cylinders' layers employed in building the lens and $k$ is an adjustable parameter used to fit $x_f$ obtained from Eq. \eqref{eq:FocalDistanceExt} to the data. The values obtained are $k =$ -0.41cm for 3.5 kHz, and $k =$ -0.28cm for 4.5 kHz, which means that the lens surface is nearer to the lens units than that usually employed in solid state physics\cite{ShuPRB93}, where the surface position is taken at a distance above the atoms equal to a half of the lattice parameter; i.e.. $k =$0 in our model.
\par
\begin{figure}
\includegraphics[width=75mm]{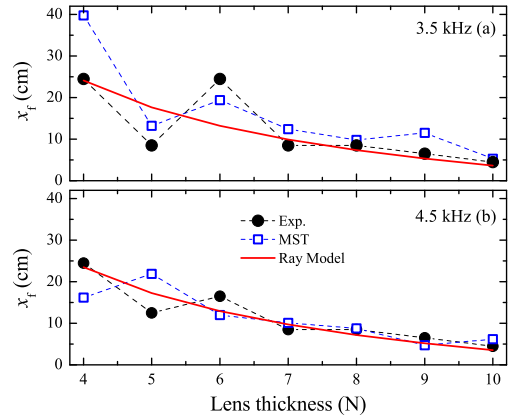}
\caption{\label{fig:FigFP}Focal spot, $x_f$, as a function of $N$, the number of layers defining the lens thickness, for frequencies: (a) 3.5kHz and (b) 4.5kHz. Black dots and hollow squares represent experimental data are results obtained from a multiple scattering theory (MST), respectively. The dashed lines between symbols are guides for the eye. Red lines represent predictions obtained by a ray model (see text). }. 
\end{figure}

 
Figure \ref{fig:FigFP} reports results obtained for $x_f$ at the lower and higher frequencies analyzed: (a) 3.5kHz and (b) 4.5kHz. 

Data and multiple scattering simulations show a better agreement with the ray model when: i) the frequency operation is near the homogenization limit and ii) for the thicker lens. 
These conclusions are physically intuitive.  On the one hand, for lower frequencies (larger wavelengths) the homogenization takes place over a larger number of unit cells in the lattice and, as a consequence, the locally designed refractive index profile $n(y)$ could be partially destroyed. 
On the second hand, thicker lenses means that sound travels longer distance inside the lens and consequently the role of surfaces and diffraction effects become negligible.      

In summary, we have demonstrated the focusing properties of 2D GRIN sonic lenses made of aluminum rods. Their broadband performance has been confirmed by numerical simulations based on a multiple scattering algorithm and by an analytical model based on ray theory. 
It can be concluded that GRIN sonic lenses are a new type of refractive acoustical devices that are feasible and reliable for possible applications not only in the sonic range, as has been demonstrated here, but also for ultrasonics and even for acoustic surface waves as has been previously proposed.

\begin{acknowledgments}
Work partially supported by USA Office of Naval Research (Grant N000140910554) and by the Spanish MICIIN under contracts TEC2007-67239 and CSD2008-66.
\end{acknowledgments}


\end{document}